\documentclass[10pt,twocolumn,showpacs,preprintnumbers,amsmath,amssymb,aps,prb,longbibliography,superscriptaddress,nofootinbib]{revtex4-2}
\usepackage{array,braket,mathtools,siunitx}

\usepackage[unicode=true,
 bookmarks=true,bookmarksnumbered=true,bookmarksopen=false,
 breaklinks=true,pdfborder={0 0 0},backref=false,colorlinks=true]{hyperref}
\hypersetup{citecolor={blue},urlcolor={magenta}}

\usepackage{cleveref}
\crefname{appendix}{App.}{Apps.}
\crefname{equation}{Eq.}{Eqs.}
\crefname{figure}{Fig.}{Figs.}
\crefname{table}{Tab.}{Tabs.}
\crefname{section}{Sec.}{Secs.}

\newcommand{\Z}{\mathbb{Z}}

\newcommand{\br}{\mathbf{r}}
\newcommand{\bk}{\mathbf{k}}
\newcommand{\ba}{\mathbf{a}}
\newcommand{\bb}{\mathbf{b}}
\newcommand{\btau}{\boldsymbol\tau}
\newcommand{\bK}{\mathbf{K}}
\newcommand{\bM}{\mathbf{M}}
\newcommand{\bG}{\mathbf{G}}

\newcommand{\bR}{\mathbf{R}}
\newcommand{\bp}{\mathbf{p}}
\newcommand{\bq}{\mathbf{q}}
\newcommand{\bsigma}{\boldsymbol\sigma}
\newcommand{\bGamma}{\boldsymbol\Gamma}
\newcommand{\bhatx}{\mathbf{\hat{x}}}
\newcommand{\bhaty}{\mathbf{\hat{y}}}
\newcommand{\bhatz}{\mathbf{\hat{z}}}

\newcommand{\bzero}{\mathbf{0}}

\begin{document}
\title{Twistronics of Kekul\'e Graphene: Honeycomb and Kagome Flat Bands}
\author{Michael G. Scheer}
\author{Biao Lian}
\affiliation{Department of Physics, Princeton University, Princeton, New Jersey 08544, USA}
\date{\today}

\begin{abstract}
Kekul\'e-O order in graphene, which has recently been realized experimentally, induces Dirac electron masses on the order of $m \sim \SI{100}{\milli\electronvolt}$. We show that twisted bilayer graphene in which one or both layers have Kekul\'e-O order exhibits nontrivial flat electronic bands on honeycomb and kagome lattices. When only one layer has Kekul\'e-O order, there is a parameter regime for which the lowest four bands at charge neutrality form an isolated two-orbital honeycomb lattice model with two flat bands. The bandwidths are minimal at a magic twist angle $\theta \approx 0.7^\circ$ and Dirac mass $m \approx \SI{100}{\milli\electronvolt}$. When both layers have Kekul\'e-O order, there is a large parameter regime around $\theta\approx 1^\circ$ and $m\gtrsim \SI{100}{\milli\electronvolt}$ in which the lowest three valence and conduction bands at charge neutrality each realize isolated kagome lattice models with one flat band, while the next three valence and conduction bands are flat bands on triangular lattices. These flat band systems may provide a new platform for strongly correlated phases of matter.
\end{abstract}

\maketitle

Moir\'e systems formed by twisting and stacking two-dimensional (2D) materials often exhibit flat electronic bands. The physics in flat bands is dominated by interactions, so strongly correlated phases often appear. A paradigmatic example is twisted bilayer graphene (TBG) at the magic angle $\theta \approx 1.05^\circ$ \cite{Bistritzer2011}, which hosts flat bands with fragile topology \cite{Po2019,Ahn2019,Song2018,Song2021} and exhibits a variety of topological and interacting phases including correlated insulators, Chern insulators, and superconductors \cite{Cao2018,Cao2018a,Yankowitz2019,Sharpe2019,Lu2019,Serlin2020,Nuckolls2020,Xie2021}. Similar flat band physics has been observed in moir\'e systems of multilayer graphene \cite{Chen2019,Park2021,Hao2021} and transition metal dichalcogenides \cite{Xu2020,Zhang2020,Foutty2023,Cai2023,Zeng2023}.

An important class of flat bands consists of those arising in tight-binding models due to wavefunction interference effects \cite{Lieb1989,Mielke1991,Calugaru2022}. Examples include the flat bands in the kagome lattice one-orbital and honeycomb lattice two-orbital tight-binding models \cite{Wu2007,Bergman2008}. Recently, we showed that such flat bands may be realized in moir\'e heterobilayers of graphene and certain 2D materials with lattice constant approximately $\sqrt{3}$ times that of graphene \cite{Scheer2022,Scheer2023}. This motivates us to search for flat bands in the twistronics of \emph{Kekul\'e graphene}, which is graphene with a $\sqrt{3} \times \sqrt{3}$ distortion. Kekul\'e graphene has been experimentally realized via epitaxial growth on a copper surface \cite{Gutierrez2016}, lithium or calcium intercalation \cite{Sugawara2011,Kanetani2012,Bao2021}, or dilute lithium deposition \cite{Cheianov2009,Qu2022}. Kekul\'e orders have also been observed in graphene in a magnetic field \cite{Li2019,Liu2022} and in correlated insulator phases of TBG \cite{Nuckolls2023}.

\begin{figure}
	\centering
	\includegraphics{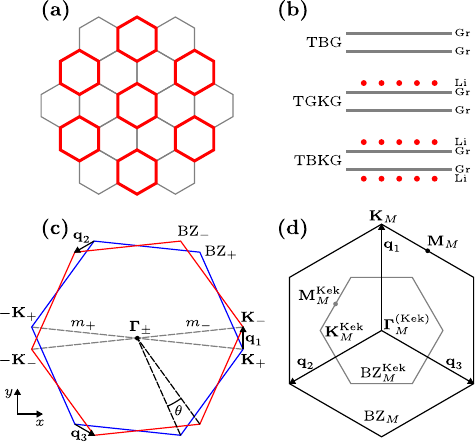}
	\caption{\textbf{(a)} Kekul\'e-O bond order in graphene. The red and gray bonds indicate hoppings of different magnitudes between neighboring on-site carbon $p_z$ orbitals. \textbf{(b)} TBG (top) and possible realizations of TGKG (middle) and TBKG (bottom) using intercalated or dilutely deposited lithium atoms (red) and graphene monolayers (gray). \textbf{(c)} The top ($l=+$) and bottom ($l=-$) layer graphene BZs before development of Kekul\'e-O order are labeled $\text{BZ}_l$. The Kekul\'e-O order induces Dirac masses $m_l$ which couple the $\bK_l$ and $-\bK_l$ points, as indicated. \textbf{(d)} The larger hexagon labeled $\text{BZ}_M$ is the moir\'e BZ for TBG and TGKG. The smaller hexagon labeled $\text{BZ}_M^{\text{Kek}}$ is the Kekul\'e moir\'e BZ for TBKG. The moir\'e $\bq_j$ vectors (defined in panel \textbf{(c)}) and high symmetry momenta in both BZs are shown.}
	\label{fig:intro}
\end{figure}

In this paper, we focus specifically on graphene with the Kekul\'e-O bond order illustrated in \cref{fig:intro}\textbf{(a)}, which can be realized by intercalation or dilute deposition of lithium \cite{Bao2021,Qu2022} and exhibits massive Dirac electrons at low energy \cite{Cheianov2009,Gamayun2018} (see Supp. \cref{app:kekule-o}). We derive a continuum model for TBG with or without Kekul\'e-O distortions and study two cases: (1) twisted graphene on Kekul\'e-O graphene (TGKG), in which only one layer has a Kekul\'e-O distortion, and (2) twisted bilayer Kekul\'e-O graphene (TBKG), in which both layers have Kekul\'e-O distortions. Possible realizations of these systems are illustrated in \cref{fig:intro}\textbf{(b)}.

Despite the Kekul\'e-O distortion, the moir\'e unit cell of TGKG is the same as that of TBG. For twist angle $\theta$ near $1^{\circ}$ and Dirac mass $m\lesssim\SI{200}{\milli\electronvolt}$ in one layer, TGKG exhibits an isolated two-orbital honeycomb lattice flat band model at charge neutrality. In particular, there is a magic angle $\theta\approx 0.7^{\circ}$ and Dirac mass $m\approx\SI{100}{\milli\electronvolt}$ for which the second valence and conduction bands become extremely flat.

In the case of TBKG, the moir\'e unit cell is enlarged to a $\sqrt{3} \times \sqrt{3}$ supercell relative to TBG when both layers have nonzero Dirac mass. For twist angle $\theta$ near $1^\circ$ and Dirac mass $m \gtrsim \SI{100}{\milli\electronvolt}$ in both layers, the lowest valence and conduction bands in TBKG at charge neutrality are one-orbital kagome lattice flat bands, with bandwidths that generally decrease with increasing $m$ and decreasing $\theta$. The next two sets of connected valence and conduction bands in this regime are flat bands on triangular lattices.

\emph{Generic continuum model.}--- We consider a twisted bilayer moir\'e system in which each layer is either graphene or Kekul\'e-O graphene. We denote the top and bottom layers by $l = +$ and $l = -$, respectively. Layer $l$ is rotated by angle $-l\theta/2$ relative to the aligned configuration, and the twist angle $\theta$ is small. We denote the lattice constant of layer $l$ by $a_l$, and define the interlayer biaxial strain $\epsilon = \ln(a_-/a_+)$. The hexagonal Brillouin zone (BZ) of graphene layer $l$ before the development of Kekul\'e-O order is shown in \cref{fig:intro}\textbf{(c)} and is denoted $\text{BZ}_l$.

The electrons at low energies in graphene layer $l$ and valley $\eta = \pm$ without Kekul\'e-O order have a Dirac Hamiltonian $h_{l,\eta}(\bp)=\hbar v_l(\eta\sigma_x p_x+\sigma_y p_y)$ at small momentum $\bp=p_x \bhatx + p_y \bhaty$ measured from $\eta \bK_l$, where the high symmetry momenta $\bK_l = \frac{4\pi}{3 a_l}R_{-l\theta/2}\bhatx$ are shown in \cref{fig:intro}\textbf{(c)}. Here, $v_l$ is the Fermi velocity of layer $l$, $\sigma_0$ is the $2 \times 2$ identity matrix, and $\sigma_x$, $\sigma_y$, and $\sigma_z$ are the Pauli matrices. The Hamiltonian $h_{l,\eta}(\bp)$ is written in the graphene sublattice basis $\alpha = +$ and $\alpha = -$, which indicate sublattices $A$ and $B$, respectively. We neglect spin degrees of freedom for simplicity.

A Kekul\'e-O order in layer $l$ modifies the Fermi velocity $v_l$ and induces an intervalley hopping term $m_l\sigma_x$ \cite{Cheianov2009,Gamayun2018} (see Supp. \cref{app:kekule-o}). This intervalley term produces a Dirac electron energy gap of $2m_l$, and we refer to $m_l$ as the Dirac mass. Additionally, we consider a potential energy difference $E_\Delta$ between the two layers, which can arise from chemical dopants or an out-of-plane displacement field.

In the continuum (i.e., small $|\bp|$) limit, we denote the real space basis for Dirac electrons at position $\br$ in layer $l$, valley $\eta$, and sublattice $\alpha$ by $\ket{\br, l, \eta, \alpha}$. The continuum Hamiltonian of this moir\'e system then takes the form $H = \int d^2\br \ket{\br} \mathcal{H}(\br) \bra{\br}$, where
\begin{widetext}
\begin{equation}\label{eq:moire-hamiltonian}
\mathcal{H}(\br) = \begin{pmatrix}
E_\Delta\sigma_0 -i\hbar v_+ \bsigma \cdot \nabla & m_+ \sigma_x & T(\br) & 0\\
m_+\sigma_x & E_\Delta\sigma_0 + i\hbar v_+ \bsigma^* \cdot \nabla & 0 & T^*(\br)\\
T^\dagger(\br) & 0 & -i\hbar v_- \bsigma \cdot \nabla & m_-\sigma_x\\
0 & T^T(\br) & m_-\sigma_x & i\hbar v_- \bsigma^* \cdot \nabla
\end{pmatrix},
\end{equation}
we have defined the basis row vector
\begin{equation}\label{eq:continuum-states}
\ket{\br} = \begin{pmatrix}
\ket{\br, +, +, +} & \ket{\br, +, +, -} & \ket{\br, +, -, +} & \ket{\br, +, -, -} & \ket{\br, -, +, +} & \ket{\br, -, +, -} & \ket{\br, -, -, +} & \ket{\br, -, -, -}
\end{pmatrix},
\end{equation}
\end{widetext}
and $\bsigma=\sigma_x \bhatx + \sigma_y \bhaty$ is the Pauli matrix vector. The interlayer moir\'e potential takes the same form as that in TBG \cite{Bistritzer2011,Song2021}, namely
\begin{equation}
\begin{split}
T(\br) &= \sum_{j=1}^3 T_{\bq_j} e^{i\bq_j \cdot \br},\quad \bq_j = R_{\zeta_j}(\bK_- - \bK_+),\\
T_{\bq_j} &= w_0 \sigma_0 + w_1 (\sigma_x\cos\zeta_j+\sigma_y\sin\zeta_j).
\end{split}
\end{equation}
Here, $\zeta_j = \frac{2\pi}{3}(j-1)$, $R_\zeta$ is the rotation matrix of angle $\zeta$, and $w_0$ and $w_1$ are the interlayer hoppings at AA and AB stacking positions, respectively. The $\bq_j$ vectors are illustrated in \cref{fig:intro}\textbf{(c)}. We have neglected the $-l\theta/2$ rotations of the Dirac Hamiltonians in \cref{eq:moire-hamiltonian}, which is a valid approximation for small $\theta$ \cite{Bistritzer2011,Song2021}. When $E_\Delta=0$ and $a_+ = a_-$, $H$ has a particle-hole symmetry $\mathcal{P} H \mathcal{P}^{-1} = -H$, where $\mathcal{P}$ is given by
\begin{equation}\label{eq:particle-hole}
\mathcal{P}\ket{\br, l, \eta, \alpha} = \eta l\ket{\mathcal{R}_\bhatx \br, l, \eta, -\alpha},
\end{equation}
and where $\mathcal{R}_\bhatx$ is the reflection matrix for the $yz$ plane. As discussed in Supp. \cref{app:symmetries}, this is different from but related to the particle-hole transformation previously discussed for TBG \cite{Song2018,Song2021}. Note that when $m_+ = m_- = E_\Delta = 0$, $v_+ = v_-$, and $a_+ = a_-$, \cref{eq:moire-hamiltonian} reduces to the two-valley model for TBG.

To a good approximation, one can neglect any changes of parameters in Kekul\'e-O graphene compared to normal graphene except for the Dirac mass $m_l$. We take $a_\pm=a_{\text{Gr}}=\SI{0.246}{\nano\meter}$ and $v_\pm = v_{\text{Gr}}$ where $\hbar v_{\text{Gr}}/a_{\text{Gr}} = \SI{2.5}{\electronvolt}$, so that the interlayer biaxial strain $\epsilon=0$. We use $w_1=\SI{110}{\milli\electronvolt}$ and $w_0/w_1=0.8$, which are typical parameters for TBG near $\theta = 1^\circ$ \cite{Bistritzer2011,Carr2019}. Additionally, we take $E_\Delta=0$ for simplicity. Results with different parameter choices including various values of $w_0/w_1$ and nonzero values of $E_\Delta$ and $\epsilon$ are given in Supp. \cref{app:parameter-variation}. Note that the sign of each Dirac mass $m_l$ in \cref{eq:moire-hamiltonian} can be flipped by applying a unitary change of basis. As a result, we take $m_l \geq 0$ without loss of generality.

\begin{figure}
	\centering
	\includegraphics{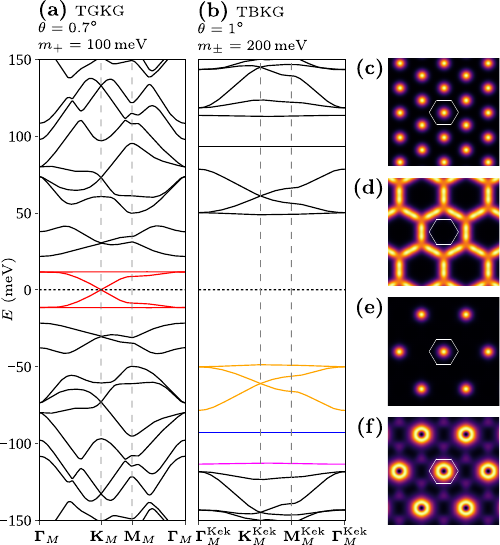}
	\caption{\textbf{(a)} Band structure of TGKG with the magic parameters $\theta = 0.7^\circ$ and $m_+ = \SI{100}{\milli\electronvolt}$. Bands $-2 \leq n \leq 2$ (shown in red) have the symmetries of a two-orbital honeycomb lattice model. \textbf{(b)} Band structure of TBKG with $\theta = 1^\circ$ and $m_\pm = \SI{200}{\milli\electronvolt}$. Bands $-3 \leq n \leq -1$ (shown in orange), $-5 \leq n \leq -4$ (shown in blue), and $n = -6$ (shown in magenta) have the symmetries of one-orbital kagome lattice, two-orbital triangular lattice, and one-orbital triangular lattice models, respectively. The dashed black lines indicate the Fermi level at charge neutrality, which must be $0$ because of the particle-hole symmetry in \cref{eq:particle-hole}. \textbf{(c)}-\textbf{(f)} The total charge density of the bands shown in red (in panel \textbf{(a)}), orange, blue, and magenta (in panel \textbf{(b)}), respectively. In each plot, the white hexagon is a unit cell for the moir\'e superlattice $L_M$ of TGKG. Note that \textbf{(d)}-\textbf{(f)} which are plots for TBKG show periodicity with respect to a $\sqrt{3} \times \sqrt{3}$ enlarged superlattice.}
	\label{fig:bands}
\end{figure}

\emph{TGKG}.--- We first consider the TGKG system illustrated in \cref{fig:intro}\textbf{(b)}, in which the top layer is Kekul\'e-O graphene with $m_+ \geq 0$ and the bottom layer is normal graphene with $m_- = 0$. In this case, the Hamiltonian commutes with the translation operators
\begin{equation}\label{eq:translation-TGKG}
T_\bR\ket{\br, l, \eta, \alpha} = e^{i(\bq_1 \cdot \bR)\eta(l-1)/2} \ket{\br + \bR, l, \eta, \alpha}
\end{equation}
for $\bR$ in the moir\'e superlattice $L_M$, which is defined as the reciprocal of the Bravais lattice $P_M$ generated by $\bq_1 - \bq_2$ and $\bq_1 - \bq_3$. As a result, TGKG has the same moir\'e unit cell as TBG. The moir\'e BZ of TGKG is the larger hexagon $\text{BZ}_M$ in \cref{fig:intro}\textbf{(d)}.

TGKG generally has magnetic space group $P61'$ (\#168.110 in the BNS setting \cite{Gallego2012}) generated by $T_\bR$ for $\bR \in L_M$, $C_{6z}$ (rotation by $\pi/3$ about $\bhatz$), and $\mathcal{T}$ (antiunitary spinless time-reversal). These operators are given in Supp. \cref{apptbl:symmetry-coreps}.

\cref{fig:bands}\textbf{(a)} shows the band structure of TGKG with $\theta = 0.7^\circ$ and $m_+ = \SI{100}{\milli\electronvolt}$. We use band index $n \neq 0$ to denote the $|n|$-th conduction (valence) band for $n > 0$ ($n < 0$). The four connected bands $-2\leq n\leq 2$ around charge neutrality (shown in red) are isolated from higher bands, and the two bands $n = \pm 2$ are extremely flat. Using magnetic topological quantum chemistry (MTQC) \cite{Bradlyn2017,Kruthoff2017,Elcoro2021}, we find that these four bands are consistent with elementary band corepresentation (EBCR) $({^1E}{^2E})_{2b}$ of $P61'$. A full table of EBCRs for each magnetic space group can be found on the Bilbao Crystallographic Server \cite{Elcoro2021,Xu2020}. EBCR $({^1E}{^2E})_{2b}$ corresponds to a system with two orbitals per site on a honeycomb lattice $L_{\text{hc}}$. These four bands can be approximately described by the honeycomb lattice tight-binding model
\begin{equation}\label{eq:two-orbital-honeycomb}
H_{\text{hc}}=\sum_{\ell,\ell'=\pm 1} t_{\ell \cdot \ell'} \sum_{\braket{j, j'} \in L_{\text{hc}}} e^{i(\ell-\ell')\varphi_{j',j}} \ket{j',\ell'}\bra{j,\ell}.
\end{equation}
Here, $t_+$ and $t_-$ are real hopping parameters, $\ket{j,\ell}$ is an orbital with angular momentum $\ell$ modulo $3$ on site $j$, $\braket{j, j'}$ runs over all nearest neighbors in $L_{\text{hc}}$, and $\varphi_{j', j}$ is the angle from an arbitrary fixed axis to the ray from site $j$ to site $j'$. When $|t_+| = |t_-|$, the highest and lowest bands of this model are exactly flat \cite{Wu2007,Calugaru2022,Scheer2023}. See Supp. \cref{app:CLS-NLS} for a construction of the compact localized states and noncontractible loop states for the flat bands in this case \cite{Rhim2019}.

Although the Wannier orbitals for bands $-2 \leq n \leq 2$ must form a honeycomb lattice, their total charge density, shown in \cref{fig:bands}\textbf{(c)}, is peaked on the triangular lattice formed by the AA stacking positions. A similar phenomenon occurs in magic angle TBG, in which case it is known that each Wannier orbital has a three-lobed ``fidget spinner" shape \cite{Koshino2018,Kang2018,Po2018}.

\begin{figure}
	\centering
	\includegraphics{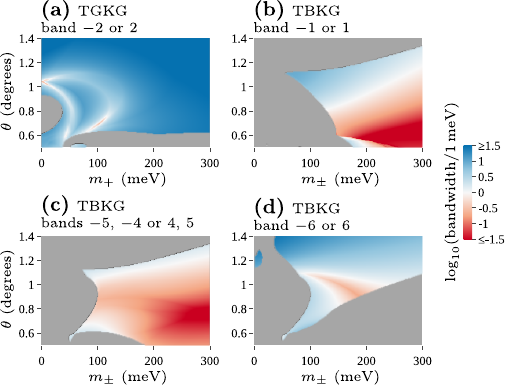}
	\caption{\textbf{(a)} The bandwidth of band $n = \pm 2$ for TGKG as a function of $\theta$ and $m_+$. \textbf{(b)}-\textbf{(d)} The bandwidth of band $n = \pm 1$, bands $-5 \leq n \leq -4$ or $4 \leq n \leq 5$, and band $n = \pm 6$ for TBKG, respectively, as a function of $\theta$ and $m_\pm$. In \textbf{(a)}-\textbf{(d)} the non-gray regions show parameters for which MTQC analysis indicates EBCR $({^1E}{^2E})_{2b}$, $(B)_{3c}$, $({^1E_2}{^2E_2})_{1a}$, and $(A)_{1a}$ of $P61'$ for TGKG bands $-2 \leq n \leq 2$, TBKG bands $-3 \leq n \leq -1$, TBKG bands $-5 \leq n \leq -4$, and TBKG band $n = -6$, respectively.}
	\label{fig:bandwidths}
\end{figure}

\cref{fig:bandwidths}\textbf{(a)} shows the bandwidth of the $n = \pm 2$ two-orbital honeycomb lattice flat bands as a function of $\theta$ and $m_+$. We observe two stripes in the parameter space in which the bandwidth reduces to approximately $\SI{1}{\milli\electronvolt}$. In particular, bandwidth minima are achieved in two regimes: near $\theta = 1^\circ$, $m_+ = 0$, which is the magic angle TBG regime, and near $\theta = 0.7^\circ$, $m_+ = \SI{100}{\milli\electronvolt}$, which we call the magic TGKG regime and illustrate in \cref{fig:bands}\textbf{(a)}. The experimentally measured Dirac masses are approximately $\SI{200}{\milli\electronvolt}$ with lithium intercalation \cite{Bao2021} and $\SI{100}{\milli\electronvolt}$ with dilute lithium deposition \cite{Qu2022}. Therefore, the magic TGKG regime for honeycomb lattice flat bands is experimentally realistic.

\emph{TBKG}.--- We now consider the TBKG system illustrated in \cref{fig:intro}\textbf{(b)}, in which both layers are Kekul\'e-O graphene. For simplicity, we assume equal Dirac masses $m_+ = m_- \geq 0$. When both $m_\pm$ are nonzero, the Hamiltonian only commutes with the translation operators $T_\bR$ in \cref{eq:translation-TGKG} for $\bR$ in the Kekul\'e moir\'e superlattice $L_M^{\text{Kek}}$, which is defined as the reciprocal of the Bravais lattice $P_M^{\text{Kek}}$ generated by $\bq_1$ and $\bq_2$. $L_M^{\text{Kek}}$ is a $\sqrt{3}\times\sqrt{3}$ superlattice of $L_M$ and the Kekul\'e moire BZ is the smaller hexagon $\text{BZ}_M^{\text{Kek}}$ in \cref{fig:intro}\textbf{(d)}.

TBKG generally has magnetic space group $P61'$ just like TGKG. In the special case considered here in which $m_+ = m_-$ and $E_\Delta = 0$, the magnetic space group expands to $P6221'$ (\#177.150 in the BNS setting \cite{Gallego2012}) because of the $C_{2x}$ (rotation by $\pi$ about the $\bhatx$) symmetry generator. However, we will use $P61'$ to emphasize that our results are stable against small $C_{2x}$ symmetry breaking perturbations. The $P61'$ symmetry operators are given in Supp. \cref{apptbl:symmetry-coreps}.

\cref{fig:bands}\textbf{(b)} shows the band structure of TBKG with $\theta = 1^\circ$ and $m_\pm = \SI{200}{\milli\electronvolt}$. There are twelve low energy bands, which result from folding the four low energy bands of TBG into $\text{BZ}_M^{\text{Kek}}$ and then adding the Dirac masses $m_\pm$. The six valence bands form three groups of connected bands. Using MTQC, we identify bands $-3 \leq n \leq -1$ (shown in orange), $-5 \leq n \leq -4$ (shown in blue), and $n = -6$ (shown in magenta) with EBCRs $(B)_{3c}$, $({^1E_2}{^2E_2})_{1a}$, and $(A)_{1a}$ of $P61'$, respectively. The conduction bands are related to the valence bands by the particle-hole transformation $\mathcal{P}$ in \cref{eq:particle-hole} and their EBCRs are given in Supp. \cref{apptbl:symmetry-coreps}. It is evident that bands $n=\pm 1, \pm 4,\pm 5,\pm 6$ are extremely flat.

We now consider the three groups of valence bands separately. First, EBCR $(B)_{3c}$ corresponds to a system with a single orbital per site on a kagome lattice $L_{\text{kag}}$. Accordingly, bands $-3 \leq n \leq -1$ can be approximately described by the kagome lattice tight-binding model
\begin{equation}\label{eq:one-orbital-kagome}
H_{\text{kag}}=t\sum_{\braket{j, j'} \in L_{\text{kag}}} \ket{j'} \bra{j},
\end{equation}
where $t$ is a real hopping parameter, $\ket{j}$ is the orbital on site $j$, and $\braket{j, j'}$ runs over all nearest neighbors in $L_{\text{kag}}$. This Hamiltonian always has an exactly flat band \cite{Bergman2008,Calugaru2022,Scheer2023} composed of compact localized states and noncontractible loop states \cite{Rhim2019}, as explained in Supp. \cref{app:CLS-NLS}. \cref{fig:bands}\textbf{(d)} shows the total charge density of bands $-3 \leq n \leq -1$, which is peaked at kagome lattice sites.

Next, EBCR $({^1E_2}{^2E_2})_{1a}$ corresponds to a system with two orbitals per site on a triangular lattice $L_{\text{tri}}$. Therefore, bands $-5 \leq n \leq -4$ can be approximately described by the triangular lattice tight-binding model
\begin{equation}
H_{\text{tri-2}} = \sum_{\ell,\ell'=\pm 1} t_{\ell \cdot \ell'} \sum_{\braket{j, j'} \in L_{\text{tri}}} e^{i(\ell-\ell')\varphi_{j',j}} \ket{j',\ell'}\bra{j,\ell}.
\end{equation}
The parameters and notations here are identical to those in \cref{eq:two-orbital-honeycomb} except that the sites $j$ here form a triangular lattice. \cref{fig:bands}\textbf{(e)} shows the total charge density of bands $-5 \leq n \leq -4$, which is peaked at triangular lattice sites.

Finally, EBCR $(A)_{1a}$ corresponds to a system with a single orbital per site on a triangular lattice $L_{\text{tri}}$. As a result, band $n = -6$ can be approximately described by the triangular lattice tight-binding model
\begin{equation}
H_{\text{tri-1}}=t\sum_{\braket{j, j'} \in L_\text{tri}} \ket{j'} \bra{j}.
\end{equation}
The parameters and notations here are identical to those in \cref{eq:one-orbital-kagome} except that the sites $j$ here form a triangular lattice. \cref{fig:bands}\textbf{(f)} shows the total charge density of band $n = -6$, which has peaks surrounding triangular lattice sites.

Figs. \ref{fig:bandwidths}\textbf{(b)}-\textbf{(d)} show the bandwidths of the $n = \pm 1$ one-orbital kagome lattice, $-5 \leq n \leq -4$ or $4 \leq n \leq 5$ two-orbital triangular lattice, and $n = \pm 6$ one-orbital triangular lattice flat bands, respectively, as a function of $\theta$ and $m_\pm$. The one-orbital kagome lattice and two-orbital triangular lattice bandwidths generally decrease with increasing $m_\pm$ and decreasing $\theta$. One the other hand, the one-orbital triangular lattice bandwidth is smallest for $0.8^\circ \lesssim \theta \lesssim 1.1^\circ$ and $\SI{100}{\milli\electronvolt} \lesssim m_\pm \lesssim \SI{200}{\milli\electronvolt}$. The Dirac masses required to realize kagome lattice and triangular lattice flat bands in TBKG are thus experimentally realistic.

\emph{Discussion}.--- We have shown that for twist angles near $\theta = 1^\circ$ and Dirac masses $m_l$ within an experimentally realistic range, TGKG exhibits a two-orbital honeycomb lattice flat band model at charge neutrality, while TBKG exhibits both kagome and triangular lattice flat band models at low energies which are quite robust against parameter variation. When interactions are included, the nontrivial flat bands of the honeycomb and kagome lattice models become promising highly tunable platforms for the realization of strongly correlated phases such as Mott insulators, charge or spin density waves, and spin liquids \cite{Yan2011,Yin2022}. We leave for future work the possibility of inducing spin-orbit coupling or magnetism through substrate coupling or other means. These effects generically produce flat (spin) Chern bands \cite{Scheer2023}, in which fractional Chern or topological insulators may be realized \cite{Tang2011}.

For TGKG, when the Dirac mass is large ($m_+ \gtrsim \SI{500}{\milli\electronvolt}$), the Kekul\'e-O graphene layer has a sufficiently large gap to allow a perturbative treatment. As explained in \cite{Scheer2023}, an effective moir\'e model can be derived for the non-Kekul\'e layer, in which the two valleys are coupled by a moir\'e potential. However, this perturbative model gives qualitatively incorrect band structures for small Dirac masses $m_+$, and in such cases the full model in \cref{eq:moire-hamiltonian} is required.

In an ideal realization of TGKG or TBKG, care must be taken to avoid charge transfer from the adatoms inducing Kekul\'e-O order into the graphene layers, since the flat bands are near the charge neutrality point of pristine graphene. In Ref. \cite{Qu2022} the authors used a dilute concentration of lithium adatoms with negligible charge transfer to the graphene layer, and observed a well-resolved Kekul\'e-O order despite the disordered adatom arrangement \cite{Cheianov2009}. Charge transfer may also be avoided through intercalation or deposition of both donor and acceptor atoms, for instance, hydrogen and lithium atoms which hole-dope and electron-dope graphene, respectively \cite{Bao2021,Qu2022}. First principles calculations and experimental studies are needed to address these issues.

\emph{Acknowledgements.}--- We thank Ali Yazdani, Christopher Guti\'errez, Changhua Bao, Jonah Herzog-Arbeitman, and Yves H. Kwan for valuable discussions. This work is supported by the Alfred P. Sloan Foundation, the National Science Foundation through Princeton University’s Materials Research Science and Engineering Center DMR-2011750, and the National Science Foundation under award DMR-2141966. Additional support is provided by the Gordon and Betty Moore Foundation through Grant GBMF8685 towards the Princeton theory program.

\bibliography{bibliography}
\clearpage
\onecolumngrid

\setcounter{equation}{0}
\setcounter{figure}{0}
\setcounter{table}{0}
\setcounter{page}{1}
\makeatletter
\renewcommand{\theequation}{S\arabic{equation}}
\renewcommand{\thefigure}{S\arabic{figure}}
\renewcommand{\thetable}{S\arabic{table}}

\begin{center}
{\bf \large Supplemental Material}
\end{center}

\section{Symmetries}\label{app:symmetries}
We now consider the crystalline symmetries of the Hamiltonian $H$ given in \cref{eq:moire-hamiltonian}. The moir\'e and Kekul\'e moir\'e reciprocal lattices $P_M$ and $P_M^{\text{Kek}}$ are Bravais lattices given by
\begin{equation}
\begin{split}
P_M &= \{n_1 \bq_1 + n_2 \bq_2 + n_3 \bq_3 | n_1, n_2, n_3 \in \Z, n_1 + n_2 + n_3 = 0\},\\
P_M^{\text{Kek}} &= \frac{1}{\sqrt{3}}R_{\pi/2}P_M = \{n_1 \bq_1 + n_2 \bq_2 + n_3 \bq_3 | n_1, n_2, n_3 \in \Z\}.
\end{split}
\end{equation}
The moir\'e and Kekul\'e moir\'e superlattices $L_M$ and $L_M^{\text{Kek}} = \sqrt{3} R_{\pi/2} L_M$ are reciprocal to $P_M$ and $P_M^{\text{Kek}}$, respectively. $H$ is always invariant under translation by elements of $L_M^{\text{Kek}}$. However, in TBG or TGKG there are also translation operators for elements in $L_M$ that commute with $H$. For a vector $\bR$, we write $T_\bR$ for the translation by $\bR$. The other relevant crystalline symmetry generators are $C_{3z}$ (rotation by $2\pi/3$ about $\bhatz$), $C_{2z}$ (rotation by $\pi$ about $\bhatz$), $C_{2x}$ (rotation by $\pi$ about $\bhatx$), and $\mathcal{T}$ (antiunitary spinless time-reversal). See \cref{apptbl:symmetry-coreps} for an enumeration of the magnetic space group corepresentations that commute with $H$ in several important cases. In each case, we give an example band corepresentation decomposition \cite{Bradlyn2017,Kruthoff2017,Elcoro2021} for the low energy bands.

The Hamiltonian for TBG (i.e., \cref{eq:moire-hamiltonian} with $m_+ = m_- = E_\Delta = 0$, $v_+ = v_-$, and $a_+ = a_-$) anticommutes with a particle-hole symmetry operator \cite{Song2018,Song2021}
\begin{equation}
\mathcal{P}_0\ket{\br, l, \eta, \alpha} = \eta l\ket{-\br, -l, \eta, \alpha}
\end{equation}
and commutes with $C_{2x}$, which takes the form
\begin{equation}
C_{2x}\ket{\br, l, \eta, \alpha} = -\ket{\mathcal{R}_\bhaty \br, -l, \eta, -\alpha}.
\end{equation}
Here, $\mathcal{R}_\bhaty$ is the reflection matrix for the $xz$ plane. Both of these symmetries are broken when $m_+ \neq m_-$, $E_\Delta \neq 0$, $v_+ \neq v_-$, or $a_+ \neq a_-$ in \cref{eq:moire-hamiltonian}. However, as long as $E_\Delta = 0$ and $a_+ = a_-$, the Hamiltonian in \cref{eq:moire-hamiltonian} anticommutes with the combined particle-hole symmetry operator $\mathcal{P} = -C_{2x}\mathcal{P}_0$, which takes the form
\begin{equation}
\mathcal{P}\ket{\br, l, \eta, \alpha} = \eta l\ket{\mathcal{R}_\bhatx \br, l, \eta, -\alpha}.
\end{equation}
Here, $\mathcal{R}_\bhatx$ is the reflection matrix for the $yz$ plane.

\begin{table*}
	\centering
	\begin{tabular}{c|c|c|c|c|c|c}
		\# & System & Constraints & $\begin{aligned} &\text{Magnetic}\\ &\text{space group}\end{aligned}$ & $\begin{aligned} &\text{Bravais}\\ &\text{lattice}\end{aligned}$ & Corep generators & Example band corep decomps\\
		\hline 1 & TBG & $\begin{aligned} &\text{Valley } \eta \text{ only}\\ & m_+ = m_- = 0\\ & v_+ = v_-\\ &E_\Delta = 0\end{aligned}$ & $\begin{aligned} &P6'2'2\\ &\text{(\#177.151)}\end{aligned}$ & $L_M$ & $\begin{aligned} T_\bR\ket{\br}_\eta &= \ket{\br + \bR}_\eta e^{-i\eta(\bq_1 \cdot \bR) \sigma_z} \otimes \sigma_0\\ C_{3z}\ket{\br}_\eta &= \ket{R_{2\pi/3}\br}_\eta \sigma_0 \otimes e^{i\eta(2\pi/3)\sigma_z}\\ C_{2x}\ket{\br} &= -\ket{\mathcal{R}_\bhaty \br} \sigma_x \otimes \sigma_x\\ C_{2z}\mathcal{T}\ket{\br}_\eta &= \ket{-\br}_\eta \sigma_0 \otimes \sigma_x\\ \end{aligned}$ & $\begin{aligned} &-1 \leq n \leq 1\text{:} \\ &(A_1)_{2c} \oplus (A_2)_{1a} \boxminus (A_1)_{1a}\end{aligned}$\\
		\hline 2 & TBG & $\begin{aligned} & m_+ = m_- = 0 \\ & v_+ = v_-\\ &E_\Delta = 0\end{aligned}$ & $\begin{aligned} &P6221'\\ &\text{(\#177.150)}\end{aligned}$ & $L_M$ & $\begin{aligned} T_\bR\ket{\br} &= \ket{\br + \bR} e^{-i(\bq_1 \cdot \bR) \sigma_z \otimes \sigma_z}\otimes \sigma_0\\ C_{3z}\ket{\br} &= \ket{R_{2\pi/3}\br} \sigma_0 \otimes e^{i(2\pi/3)\sigma_z \otimes \sigma_z}\\ C_{2z}\ket{\br} &= \ket{-\br} \sigma_0 \otimes \sigma_x \otimes \sigma_x\\ C_{2x}\ket{\br} &= -\ket{\mathcal{R}_\bhaty \br} \sigma_x \otimes \sigma_0 \otimes \sigma_x\\ \mathcal{T}\ket{\br} &= \ket{\br} \sigma_0 \otimes \sigma_x \otimes \sigma_0 \end{aligned}$ & $\begin{aligned} &-2 \leq n \leq 2\text{:}\\ &(A_2)_{2c} \oplus (A_1)_{2c}\end{aligned}$\\
		\hline 3 & TGKG & $\begin{aligned} & m_- = 0\end{aligned}$ & $\begin{aligned} &P61'\\ &\text{(\#168.110)}\end{aligned}$ & $L_M$ & $\begin{aligned} T_\bR\ket{\br} &= \ket{\br + \bR}(\sigma_0 \otimes \sigma_0)\oplus (e^{-i(\bq_1 \cdot \bR) \sigma_z} \otimes \sigma_0)\\ C_{3z}\ket{\br} &= \ket{R_{2\pi/3}\br} \sigma_0 \otimes e^{i(2\pi/3)\sigma_z \otimes \sigma_z}\\ C_{2z}\ket{\br} &= \ket{-\br} \sigma_0 \otimes \sigma_x \otimes \sigma_x\\ \mathcal{T}\ket{\br} &= \ket{\br} \sigma_0 \otimes \sigma_x \otimes \sigma_0 \end{aligned}$ & $\begin{aligned} &-2\leq n \leq 2\text{:}\\ & ({^1E}{^2E})_{2b}\end{aligned}$\\
		\hline 4 & TGKG & $\begin{aligned} & m_+ = 0 \end{aligned}$ & $\begin{aligned} &P61'\\ &\text{(\#168.110)}\end{aligned}$ & $L_M$ & $\begin{aligned} T_\bR\ket{\br} &= \ket{\br + \bR}(e^{i(\bq_1 \cdot \bR) \sigma_z} \otimes \sigma_0) \oplus (\sigma_0 \otimes \sigma_0)\\ C_{3z}\ket{\br} &= \ket{R_{2\pi/3}\br} \sigma_0 \otimes e^{i(2\pi/3)\sigma_z \otimes \sigma_z}\\ C_{2z}\ket{\br} &= \ket{-\br} \sigma_0 \otimes \sigma_x \otimes \sigma_x\\ \mathcal{T}\ket{\br} &= \ket{\br} \sigma_0 \otimes \sigma_x \otimes \sigma_0 \end{aligned}$ & $\begin{aligned} &-2\leq n \leq 2\text{:}\\ &({^1E}{^2E})_{2b}\end{aligned}$\\
		\hline 5 & TBKG & N/A & $\begin{aligned} &P61'\\ &\text{(\#168.110)}\end{aligned}$ & $L_M^{\text{Kek}}$ & $\begin{aligned} T_\bR\ket{\br} &= \ket{\br + \bR}\\ C_{3z}\ket{\br} &= \ket{R_{2\pi/3}\br} \sigma_0 \otimes e^{i(2\pi/3)\sigma_z \otimes \sigma_z}\\ C_{2z}\ket{\br} &= \ket{-\br} \sigma_0 \otimes \sigma_x \otimes \sigma_x\\ \mathcal{T}\ket{\br} &= \ket{\br} \sigma_0 \otimes \sigma_x \otimes \sigma_0 \end{aligned}$ & $\begin{aligned} n = 6\text{: } &(B)_{1a}\\ 4\leq n \leq 5\text{: } &({^1E_1}{^2E_1})_{1a}\\ 1 \leq n \leq 3\text{: } &(A)_{3c}\\ -3\leq n \leq -1\text{: } &(B)_{3c}\\ -5\leq n \leq -4\text{: } &({^1E_2}{^2E_2})_{1a}\\ n = -6\text{: } &(A)_{1a}\end{aligned}$
	\end{tabular}
	\caption{Magnetic topological quantum chemistry \cite{Bradlyn2017,Kruthoff2017,Elcoro2021} analysis for the model in \cref{eq:moire-hamiltonian} in several cases. Row $1$ describes a one-valley model while rows $2$ through $5$ describe two-valley models. The basis vector $\ket{\br}$ is defined in \cref{eq:continuum-states} so that the tensor product factors for all two-valley models are layer, valley, and sublattice, in that order. For the one-valley model, we use $\ket{\br}_\eta = \begin{pmatrix}
	\ket{\br, +, \eta, +} & \ket{\br, +, \eta, -} & \ket{\br, -, \eta, +} & \ket{\br, -, \eta, -}
	\end{pmatrix}$ so that the tensor product factors are layer and sublattice, in that order. $R_{2\pi/3}$ is the rotation matrix of angle $2\pi/3$ and $R_\bhaty$ is the reflection matrix for $xz$ plane. The numbers for the magnetic space groups are given in the BNS setting \cite{Gallego2012} and elementary band corepresentation tables for each magnetic space group can be found on the Bilbao Crystallographic Server \cite{Elcoro2021,Xu2020a}. For each band corepresentation decomposition, we use $w_1 = \SI{110}{\milli\electronvolt}$, $w_0/w_1 = 0.8$, $E_\Delta = 0$, $a_\pm = a_{\text{Gr}} = \SI{0.246}{\nano\meter}$, and $v_\pm = v_{\text{Gr}}$, where $\hbar v_{\text{Gr}}/a_{\text{Gr}} = \SI{2.5}{\electronvolt}$. The other parameters are as follows. In rows $1$ and $2$, we use $\theta = 1^\circ$ and $m_\pm = 0$. In row $3$, we use $\theta = 0.7^\circ$, $m_+ = \SI{100}{\milli\electronvolt}$, and $m_- = 0$, as in \cref{fig:bands}\textbf{(a)}. In row 4, we use $\theta = 0.7^\circ$, $m_+ = 0$, and $m_- = \SI{100}{\milli\electronvolt}$. In row 5, we use $\theta = 1^\circ$, and $m_+ = m_- = \SI{200}{\milli\electronvolt}$, as in \cref{fig:bands}\textbf{(b)}.}
	\label{apptbl:symmetry-coreps}
\end{table*}

\section{Kekul\'e-O graphene}\label{app:kekule-o}
\begin{figure}[h]
	\centering
	\includegraphics{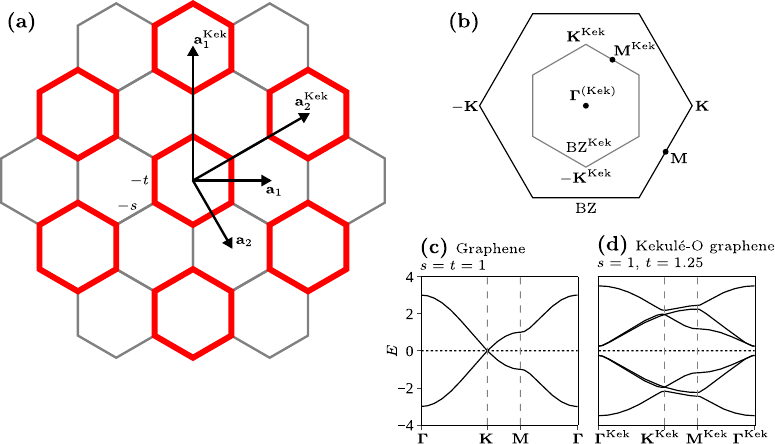}
	\caption{\textbf{(a)} Illustration of the lattice vectors and hoppings for Kekul\'e-O bond order in graphene. \textbf{(b)} The graphene BZ (black), Kekul\'e BZ (gray), and their high symmetry momenta. \textbf{(c)} Example gapless band structure of graphene using the Hamiltonian in \cref{appeq:graphene-hamiltonian} with $t = 1$. \textbf{(d)} Example band structure of Kekul\'e-O graphene using the Hamiltonian in \cref{appeq:kekule-o-hamiltonian} with $s = 1$, $t = 1.25$. There is an energy gap of $2|t-s| = 0.5$.}
	\label{appfig:kekule-o}
\end{figure}

In the following subsections, we derive and analyze tight-binding models and low energy continuum models for graphene both with and without a Kekul\'e-O distortion. As illustrated in \cref{appfig:kekule-o}\textbf{(a)}, we consider a nearest neighbor tight-binding model with a single $p_z$ orbital on each site of a honeycomb lattice. For simplicity, we neglect spin degrees of freedom. The nearest neighbor bonds shown with thick red lines (thin gray lines) have real hopping parameter $-t$ ($-s$).

We denote by $L$ the triangular Bravais lattice formed by the centers of all the hexagons. Similarly, we define $L^{\text{Kek}} = \sqrt{3}R_{\pi/2}L$ to be the Bravais lattice formed by the centers of all the hexagons bounded by thick red lines. The primitive vectors $\ba_1$ and $\ba_2$ ($\ba^{\text{Kek}}_1$ and $\ba^{\text{Kek}}_2$) for $L$ ($L^{\text{Kek}}$) are shown in \cref{appfig:kekule-o}\textbf{(a)}. These primitive vectors are given by
\begin{equation}\label{appeq:define-a-vectors}
\ba_1 = a \bhatx,\quad \ba_2 = R_{-\pi/3} \ba_1, \quad \ba^{\text{Kek}}_1 = \sqrt{3}R_{\pi/2}\ba_1,\quad \ba^{\text{Kek}}_2 = \sqrt{3}R_{\pi/2}\ba_2
\end{equation}
for a positive constant $a$.

We denote the lattices reciprocal to $L$ and $L^{\text{Kek}}$ by $P$ and $P^{\text{Kek}} = R_{\pi/2}P/\sqrt{3}$, respectively. The primitive vectors $\bb_1$ and $\bb_2$ ($\bb^{\text{Kek}}_1$ and $\bb^{\text{Kek}}_2$) for $P$ ($P^{\text{Kek}}$) are given by
\begin{equation}
\bb_1 = R_{2\pi/3}\bb_2,\quad \bb_2 = -4\pi\bhaty/(a\sqrt{3}), \quad \bb^{\text{Kek}}_1 = R_{\pi/2}\bb_1/\sqrt{3},\quad \bb^{\text{Kek}}_2 = R_{\pi/2}\bb_2/\sqrt{3}.
\end{equation}
We define high symmetry crystal momenta
\begin{equation}
\bGamma = \bzero,\quad \bK = \frac{2}{3}\bb_1 + \frac{1}{3}\bb_2,\quad \bM = \frac{1}{2}\bb_1 + \frac{1}{2}\bb_2,\quad \bGamma^{\text{Kek}} = \bzero,\quad \bK^{\text{Kek}} = R_{\pi/2}\bK/\sqrt{3},\quad \bM^{\text{Kek}} = R_{\pi/2}\bM/\sqrt{3}.
\end{equation}
The Brillouin zones $\text{BZ}$ and $\text{BZ}^{\text{Kek}}$ and the high symmetry crystal momenta are illustrated in \cref{appfig:kekule-o}\textbf{(b)}. Note that
\begin{equation}\label{appeq:define-K-explicitly}
\bK =4\pi \bhatx/(3a) = \bb^{\text{Kek}}_2
\end{equation}
so that $\bGamma$, $\bK$, and $-\bK$ are all elements of $P^{\text{Kek}}$.

\subsection{Without Kekul\'e-O distortion}\label{app:without-kekule}
We first consider the special case in which $s = t$ so that the model is symmetric under translation by elements of $L$ and the system has lattice constant $a$. We enumerate the lattice sites as $\br + \alpha \btau$ for $\br \in L$ and $\alpha \in \{+, -\}$, where
\begin{equation}\label{appeq:define-tau}
\btau = R_{\pi/2}\ba_1/\sqrt{3}.
\end{equation}
The $A$ and $B$ sublattices correspond here to sites with $\alpha = +$ and $\alpha = -$, respectively. We denote the orbital at site $\br + \alpha\btau$ by $\ket{\br, \alpha}$ and define Bloch states
\begin{equation}\label{appeq:define-graphene-bloch-states}
\ket{\bk, \alpha} = \frac{1}{\sqrt{|\text{BZ}|}} \sum_{\br \in L} e^{i\bk \cdot (\br + \alpha \btau)} \ket{\br, \alpha}.
\end{equation}
These Bloch states carry a corepresentation of the magnetic space group $P6mm1'$ (\#183.186 in the BNS setting \cite{Gallego2012}) determined by
\begin{equation}\label{appeq:graphene-corep}
\begin{split}
T_{\bR} \ket{\bk, \alpha} &= e^{-i\bk \cdot \bR}\ket{\bk, \alpha},\\
C_{3z} \ket{\bk, \alpha} &= \ket{R_{2\pi/3} \bk, \alpha},\\
M_x \ket{\bk, \alpha} &= \ket{\mathcal{R}_\bhatx \bk, \alpha},\\
M_y \ket{\bk, \alpha} &= \ket{\mathcal{R}_\bhaty \bk, -\alpha},\\
\mathcal{T} \ket{\bk, \alpha} &= \ket{-\bk, \alpha},
\end{split}
\end{equation}
where $T_\bR$ denotes translation by $\bR \in L$, $C_{3z}$ denotes rotation by $2\pi/3$ about $\bhatz$, $M_x$ denotes reflection through the $yz$ plane, $M_y$ denotes reflection through the $xz$ plane, and $\mathcal{T}$ denotes antilinear spinless time-reversal. One could enlarge the group by considering also three dimensional inversion symmetry, but for simplicity we will not do so. The Hamiltonian commutes with the operators in \cref{appeq:graphene-corep} and takes the form
\begin{equation}\label{appeq:graphene-hamiltonian}
H \ket{\bk, \alpha} = -t \sum_{j=1}^3 e^{-i\alpha \bk \cdot R_{\zeta_j}\btau} \ket{\bk, -\alpha}.
\end{equation}
Equivalently, for a fixed $\bk$ the Hamiltonian acts as the matrix
\begin{equation}
-t\sum_{j=1}^3 \begin{pmatrix}
0 & e^{i\bk \cdot R_{\zeta_j}\btau}\\
e^{-i\bk \cdot R_{\zeta_j}\btau} & 0
\end{pmatrix}
\end{equation}
in the basis $\ket{\bk, +}$, $\ket{\bk, -}$.

As shown in \cref{appfig:kekule-o}\textbf{(c)}, the low energy physics for the Hamiltonian in \cref{appeq:graphene-hamiltonian} consists of Dirac cones at the $\bK$ and $-\bK$ points. To investigate this analytically, we take $\bk = \eta \bK + \bp$ for $\eta \in \{+, -\}$ and expand $H\ket{\bk, \alpha}$ to first order in $|\bp|$. Using the identity
\begin{equation}\label{appeq:K-dot-R-identity}
\bK \cdot R_{\zeta_j} \btau = -\zeta_j
\end{equation}
for $j \in \{1, 2, 3\}$, we have
\begin{equation}\label{appeq:approx-graphene-K}
\begin{split}
H \ket{\eta \bK + \bp, \alpha} &= -t\sum_{j=1}^3 e^{i\eta\alpha\zeta_j}e^{-i\alpha\bp \cdot R_{\zeta_j}\btau} \ket{\eta \bK + \bp, -\alpha}\\
&\approx -t\sum_{j=1}^3 e^{i\eta\alpha\zeta_j}(1 -i\alpha\bp \cdot R_{\zeta_j}\btau) \ket{\eta \bK + \bp, -\alpha}\\
&= \bp \cdot \left(\frac{i\alpha a t}{\sqrt{3}}\sum_{j=1}^3 e^{i\eta\alpha\zeta_j} R_{\zeta_j}\bhaty\right)\ket{\eta \bK + \bp, -\alpha}\\
&= \hbar v_F \bp \cdot (\eta \bhatx + i\alpha\bhaty)\ket{\eta \bK + \bp, -\alpha}
\end{split}
\end{equation}
where
\begin{equation}\label{appeq:define-v_F}
v_F = \frac{a t \sqrt{3}}{2\hbar}.
\end{equation}
Equivalently, for a fixed $\eta$ and $\bp$ the Hamiltonian acts as the matrix
\begin{equation}\label{appeq:explicit-Dirac-cone}
\hbar v_F (\eta \sigma_x \bhatx + \sigma_y \bhaty) \cdot \bp
\end{equation}
in the basis $\ket{\eta \bK + \bp, +}$, $\ket{\eta \bK + \bp, -}$, to first order in $|\bp|$. We recognize \cref{appeq:explicit-Dirac-cone} as a Dirac cone with Fermi velocity $v_F$.

It will also be useful in the next section to expand $H\ket{\bGamma + \bp, \alpha}$ to first order in $|\bp|$. This is simply
\begin{equation}\label{appeq:approx-graphene-Gamma}
\begin{split}
H \ket{\bGamma + \bp, \alpha} &= -t \sum_{j=1}^3 e^{-i\alpha \bp \cdot R_{\zeta_j}\btau}\ket{\bGamma + \bp, -\alpha}\\
&\approx -t \sum_{j=1}^3 (1-i\alpha \bp \cdot R_{\zeta_j}\btau)\ket{\bGamma + \bp, -\alpha}\\
&= -3t\ket{\bGamma + \bp, -\alpha}.
\end{split}
\end{equation}

\subsection{With Kekul\'e-O distortion}
We now consider the general case in which $s$ and $t$ may be different, so we are only guaranteed that the model is symmetric under translation by elements of $L^{\text{Kek}}$. When $s \neq t$, the system has lattice constant $a\sqrt{3}$. It is convenient to enumerate the lattice sites as $\br + \alpha R_{\zeta_j} \btau$ for $\br \in L^{\text{Kek}}$, $\alpha \in \{+, -\}$, and $j \in \{1, 2, 3\}$, where $\btau$ is given by \cref{appeq:define-tau}. As in \cref{app:without-kekule}, the $A$ and $B$ sublattices are defined by $\alpha = +$ and $\alpha = -$, respectively. We denote the orbital at site $\br + \alpha R_{\zeta_j}\btau$ by $\ket{\br, \alpha, j}$ and define Bloch states
\begin{equation}\label{appeq:define-kekule-o-bloch-states}
\ket{\bk, \alpha, j} = \frac{1}{\sqrt{\left|\text{BZ}^{\text{Kek}}\right|}} \sum_{\br \in L^{\text{Kek}}} e^{i\bk \cdot \left(\br + \alpha R_{\zeta_j}\btau\right)}\ket{\br, \alpha, j}.
\end{equation}
These Bloch states carry a corepresentation of $P6mm1'$ determined by
\begin{equation}\label{appeq:kekule-o-corep}
\begin{split}
T_\bR \ket{\bk, \alpha, j} &= e^{-i\bk \cdot \bR}\ket{\bk, \alpha, j},\\
C_{3z} \ket{\bk, \alpha, j} &= \ket{R_{2\pi/3}\bk, \alpha, j + 1},\\
M_x \ket{\bk, \alpha, j} &= \ket{\mathcal{R}_\bhatx \bk, \alpha, 2-j},\\
M_y \ket{\bk, \alpha, j} &= \ket{\mathcal{R}_\bhaty \bk, -\alpha, 2-j},\\
\mathcal{T} \ket{\bk, \alpha, j} &= \ket{-\bk, \alpha, j},
\end{split}
\end{equation}
where $T_\bR$ now denotes translation by $\bR \in L^{\text{Kek}}$, and the $j$ indices are defined modulo $3$. The Hamiltonian commutes with the operators in \cref{appeq:kekule-o-corep} and takes the form
\begin{equation}\label{appeq:kekule-o-hamiltonian}
H^{\text{Kek}} \ket{\bk, \alpha, j} = -se^{-i\alpha \bk \cdot R_{\zeta_j}\btau} \ket{\bk, -\alpha, j} -t\sum_{j'\neq j} e^{i\alpha\bk \cdot \left(R_{\zeta_j} + R_{\zeta_{j'}}\right)\btau} \ket{\bk, -\alpha, j'}.
\end{equation}
As illustrated in \cref{appfig:kekule-o}\textbf{(a)}, the terms proportional to $-s$ and $-t$ produce hopping between and within the hexagons bounded by thick red lines, respectively.

As shown in \cref{appfig:kekule-o}\textbf{(d)}, when $s$ differs slightly from $t$ the low energy physics for the Hamiltonian in \cref{appeq:kekule-o-hamiltonian} consists of two gapped Dirac cones at the $\bGamma^{\text{Kek}}$ point. To investigate this analytically, we first note that the Bloch states in \cref{appeq:define-graphene-bloch-states,appeq:define-kekule-o-bloch-states} are related by
\begin{equation}
\ket{\bk, \alpha} = \frac{1}{\sqrt{3}}\sum_{j=1}^3 \ket{\bk, \alpha, j}
\end{equation}
and additionally the Bloch states in \cref{appeq:define-kekule-o-bloch-states} satisfy
\begin{equation}
\ket{\bk + \bG, \alpha, j} = e^{i\alpha\bG \cdot R_{\zeta_j}\btau}\ket{\bk, \alpha, j} 
\end{equation}
for $\bG \in P^{\text{Kek}}$. Recalling that $\eta \bK \in P^{\text{Kek}}$ for $\eta \in \{+,0,-\}$ and using \cref{appeq:K-dot-R-identity}, we then have
\begin{equation}
\ket{\eta \bK + \bp, \alpha} = \frac{1}{\sqrt{3}} \sum_{j=1}^3 e^{-i\eta\alpha \zeta_j}\ket{\bGamma^{\text{Kek}} + \bp, \alpha, j},
\end{equation}
so that the states $\ket{\eta \bK + \bp, \alpha}$ and $\ket{\bGamma^{\text{Kek}} + \bp, \alpha, j}$ for $\eta \in \{+, 0, -\}$ and $j \in \{1, 2, 3\}$ are related by a unitary change of basis. We can therefore study the low energy physics of $H^{\text{Kek}}$ near the $\bGamma_{\text{Kek}}$ point by expanding $H^{\text{Kek}}\ket{\eta \bK + \bp, \alpha}$ to first order in $|\bp|$. Using the fact that the Hamiltonians in \cref{appeq:graphene-hamiltonian,appeq:kekule-o-hamiltonian} coincide when $s = t$, we have
\begin{equation}
\begin{split}
(H^{\text{Kek}} - H)\ket{\eta \bK + \bp, \alpha} &= \frac{t-s}{\sqrt{3}}\sum_{j=1}^3 e^{-i\eta\alpha\zeta_j} e^{-i\alpha \bp \cdot R_{\zeta_j}\btau}\ket{\bGamma^{\text{Kek}} + \bp, -\alpha, j}\\
&= \frac{t-s}{3}\sum_{\eta'\in\{+,0,-\}}\sum_{j=1}^3 e^{-i(\eta' + \eta)\alpha\zeta_j} e^{-i\alpha \bp \cdot R_{\zeta_j}\btau}\ket{\eta'\bK + \bp, -\alpha}
\end{split}
\end{equation}
where $H$ is given by \cref{appeq:graphene-hamiltonian}. Using \cref{appeq:approx-graphene-K,appeq:approx-graphene-Gamma}, for $\eta \neq 0$ we find
\begin{equation}
H^{\text{Kek}}\ket{\eta \bK + \bp, \alpha} \approx \hbar v_F^{\text{Kek}} \bp \cdot (\eta\bhatx + i\alpha \bhaty)\ket{\eta \bK+ \bp, -\alpha} + (t-s)\ket{-\eta \bK + \bp, -\alpha} + \hbar v_0 \bp \cdot (-\eta \bhatx + i\alpha \bhaty) \ket{\bGamma + \bp, -\alpha}
\end{equation}
to first order in $|\bp|$, where
\begin{equation}
v_F^{\text{Kek}} = \frac{s + 2t}{3} \frac{a\sqrt{3}}{2\hbar},\quad v_0 = \frac{s-t}{3}\frac{a\sqrt{3}}{2\hbar}.
\end{equation}
Similarly, for $\eta = 0$ we find
\begin{equation}
H^{\text{Kek}}\ket{\bGamma + \bp, \alpha} \approx -(s + 2t) \ket{\bGamma + \bp, -\alpha} + \sum_{\eta = \pm} \hbar v_0 \bp \cdot (-\eta\bhatx + i\alpha \bhaty) \ket{\eta \bK + \bp, -\alpha}
\end{equation}
to first order in $|\bp|$.

\subsection{Low energy continuum models}
We assume that $|v_0/v_F^{\text{Kek}}|$ is small enough for us to treat perturbatively the terms in $H^{\text{Kek}}$ coupling states near $\pm\bK$ to states near $\bGamma$. Since these coupling terms are proportional to $|\bp|$, they make no contribution to the low energy physics to first order in $|\bp|$. The low energy physics for $H^{\text{Kek}}$ near $\bGamma^{\text{Kek}}$ is therefore described by the effective continuum Hamiltonian
\begin{equation}
H^{\text{Kek}}_0\ket{\bp, \eta, \alpha} = \hbar v_F^{\text{Kek}} \bp \cdot (\eta\bhatx + i\alpha \bhaty)\ket{\bp, \eta, -\alpha} + m\ket{\bp, -\eta, -\alpha}
\end{equation}
where
\begin{equation}
m = t - s
\end{equation}
and $\ket{\bp, \eta, \alpha}$ for $\eta, \alpha \in \{+, -\}$ is a set of continuum states normalized by
\begin{equation}
\braket{\bp', \eta', \alpha' | \bp, \eta, \alpha} = \delta^2(\bp'-\bp)\delta_{\eta',\eta}\delta_{\alpha',\alpha}.
\end{equation}
Defining the row vector of states
\begin{equation}\label{appeq:define-ket-p}
\ket{\bp} = \begin{pmatrix}
\ket{\bp, +, +} & \ket{\bp, +, -} & \ket{\bp, -, +} & \ket{\bp, -, -}
\end{pmatrix}
\end{equation}
we can write
\begin{equation}\label{appeq:kekule-o-continuum-hamiltonian-momentum}
\begin{split}
H^{\text{Kek}}_0 &= \int d^2\bp \ket{\bp} \mathcal{H}^{\text{Kek}}(\bp) \bra{\bp}\\
\mathcal{H}^{\text{Kek}}(\bp) &= \begin{pmatrix}
\hbar v_F^{\text{Kek}} \bsigma \cdot \bp & m\sigma_x\\
m\sigma_x & -\hbar v_F^{\text{Kek}} \bsigma^* \cdot \bp
\end{pmatrix}.
\end{split}
\end{equation}

In order to describe $H^{\text{Kek}}_0$ in real space, we define real space continuum states
\begin{equation}
\ket{\br, \eta, \alpha} = \frac{1}{2\pi}\int d^2\bp e^{-i\bp \cdot \br}\ket{\bp, \eta, \alpha}
\end{equation}
which satisfy
\begin{equation}
\braket{\br', \eta', \alpha' | \br, \eta, \alpha} = \delta^2(\br'-\br)\delta_{\eta',\eta}\delta_{\alpha',\alpha}.
\end{equation}
Defining the row vector of states
\begin{equation}
\ket{\br} = \begin{pmatrix}
\ket{\br, +, +} & \ket{\br, +, -} & \ket{\br, -, +} & \ket{\br, -, -}
\end{pmatrix}
\end{equation}
we can write
\begin{equation}\label{appeq:kekule-o-continuum-hamiltonian-real}
\begin{split}
H^{\text{Kek}}_0 &= \int d^2\br \ket{\br} \mathcal{H}^{\text{Kek}}(\br) \bra{\br}\\
\mathcal{H}^{\text{Kek}}(\br) &= \begin{pmatrix}
-i\hbar v_F^{\text{Kek}} \bsigma \cdot \nabla & m\sigma_x\\
m\sigma_x & i\hbar v_F^{\text{Kek}} \bsigma^* \cdot \nabla
\end{pmatrix}.
\end{split}
\end{equation}
In the special case that $s = t$, we have $m = 0$ and $v_F^{\text{Kek}} = v_F$ so that $H^{\text{Kek}}_0$ simplifies to the continuum model $H^{\text{Gr}}_0$ for graphene given by
\begin{equation}\label{appeq:graphene-continuum-hamiltonian-real}
\begin{split}
H^{\text{Gr}}_0 &= \int d^2\br \ket{\br} \mathcal{H}^{\text{Gr}}(\br) \bra{\br}\\
\mathcal{H}^{\text{Gr}}(\br) &= \begin{pmatrix}
-i\hbar v_F \bsigma \cdot \nabla & 0\\
0 & i\hbar v_F \bsigma^* \cdot \nabla
\end{pmatrix}.
\end{split}
\end{equation}

\subsection{Spectrum and symmetry properties}
We now consider the spectrum and symmetry properties of the continuum Hamiltonian $H^{\text{Kek}}_0$ given by \cref{appeq:kekule-o-continuum-hamiltonian-momentum}. We first note that
\begin{equation}
U^\dagger_0\mathcal{H}^{\text{Kek}}(\bp) U_0 = \begin{pmatrix}
\hbar v_F^{\text{Kek}} \bsigma \cdot \bp + m \sigma_z & 0\\
0 & -\hbar v_F^{\text{Kek}} \bsigma^* \cdot \bp + m \sigma_z
\end{pmatrix}
\end{equation}
where
\begin{equation}
U_0 = \frac{1}{\sqrt{2}}\begin{pmatrix}
\sigma_0 & -i\sigma_y\\
-i\sigma_y & \sigma_0
\end{pmatrix}
\end{equation}
is a unitary matrix. As a result, the spectrum of $H^{\text{Kek}}_0$ consists of two bands with dispersion $E_+(\bp)$ and two bands with dispersion $E_-(\bp)$ where
\begin{equation}
E_\pm(\bp) = \pm \sqrt{(\hbar v_F^{\text{Kek}} |\bp|)^2 + m^2}.
\end{equation}
In particular, there is an energy gap of $2|m|$, as can be seen in \cref{appfig:kekule-o}\textbf{(d)}.

Next, we note that $H^{\text{Kek}}_0$ commutes with a corepresentation of the magnetic point group $6mm1'$ (see Appendix H of Ref. \cite{Scheer2023}) determined by
\begin{equation}\label{appeq:kekule-o-continuum-corep}
\begin{split}
C_{3z} \ket{\bp} &= \ket{R_{2\pi/3}\bp} e^{i(2\pi/3)\sigma_z \otimes \sigma_z},\\
M_x \ket{\bp} &= \ket{\mathcal{R}_\bhatx \bp} \sigma_x \otimes \sigma_0,\\
M_y \ket{\bp} &= \ket{\mathcal{R}_\bhaty \bp} \sigma_0 \otimes \sigma_x,\\
\mathcal{T}\ket{\bp} &= \ket{-\bp} \sigma_x \otimes \sigma_0.
\end{split}
\end{equation}
In accordance with \cref{appeq:define-ket-p}, the first and second tensor product factors here indicate valley and sublattice, respectively. The corepresentation in \cref{appeq:kekule-o-continuum-corep} can be derived from \cref{appeq:graphene-corep} under the identification of $\ket{\bp, \eta, \alpha}$ with $\ket{\eta \bK + \bp, \alpha}$. To simplify this corepresentation, we take unitary change of basis
\begin{equation}\label{appeq:define-U}
\ket{\bp}' = \ket{\bp} U_1, \quad U_1 = \frac{1}{\sqrt{2}}\begin{pmatrix}
\sigma_x & -i\sigma_x\\
\sigma_0 & i\sigma_0
\end{pmatrix}
\end{equation}
which mixes the valley and sublattice degrees of freedom. In this basis, the operators in \cref{appeq:kekule-o-continuum-corep} become
\begin{equation}\label{appeq:kekule-o-continuum-corep-diagonalized}
\begin{split}
C_{3z}\ket{\bp}' &= \ket{R_{2\pi/3}\bp}' e^{-i(2\pi/3)\sigma_z} \oplus e^{-i(2\pi/3)\sigma_z},\\
M_x \ket{\bp}' &= \ket{\mathcal{R}_\bhatx \bp}' \sigma_x \oplus (-\sigma_x),\\
M_y \ket{\bp}' &= \ket{\mathcal{R}_\bhaty \bp}' \sigma_x \oplus \sigma_x,\\
\mathcal{T}\ket{\bp}' &= \ket{-\bp}' \sigma_x \oplus \sigma_x.
\end{split}
\end{equation}
At $\bp = \bzero$, this corepresentation decomposes as the direct sum of two different 2D irreducible corepresentations which are called $E_2$ and $E_1$ (see Table III of Ref. \cite{Scheer2023}). In this basis, the Hamiltonian matrix becomes
\begin{equation}\label{appeq:kekule-o-continuum-hamiltonian-diagonalized}
U_1^\dagger \mathcal{H}^{\text{Kek}}(\bp) U_1 = \begin{pmatrix}
m\sigma_0 & -i\hbar v_F^{\text{Kek}} \bsigma^* \cdot \bp\\
i\hbar v_F^{\text{Kek}}\bsigma^* \cdot \bp & -m\sigma_0
\end{pmatrix}
\end{equation}
which is diagonal when $\bp = \bzero$.

\section{Compact localized states and noncontractible loop states}\label{app:CLS-NLS}
The tight-binding Hamiltonians in \cref{eq:two-orbital-honeycomb,eq:one-orbital-kagome} have exactly flat bands with symmetry protected band touchings at the $\bGamma$ point \cite{Wu2007,Bergman2008,Calugaru2022,Scheer2023}. As explained in Ref. \cite{Rhim2019}, the flat bands of such models can generally be spanned by a collection of compact localized states (CLSs) and noncontractible loop states (NLSs), all of which are eigenstates of the Hamiltonian with a fixed energy. As the names suggest, CLSs have compact support, while NLSs are compact in one direction, but span a noncontractible loop in the other direction, when the system is defined with periodic boundary conditions. In the following subsections, we construct the CLSs and NLSs for the flat bands in the Hamiltonians in \cref{eq:two-orbital-honeycomb,eq:one-orbital-kagome}. In both cases, we assume the model is defined with periodic boundary conditions and a total of $N_{\text{tot}}$ unit cells.

\begin{figure}
	\centering
	\includegraphics{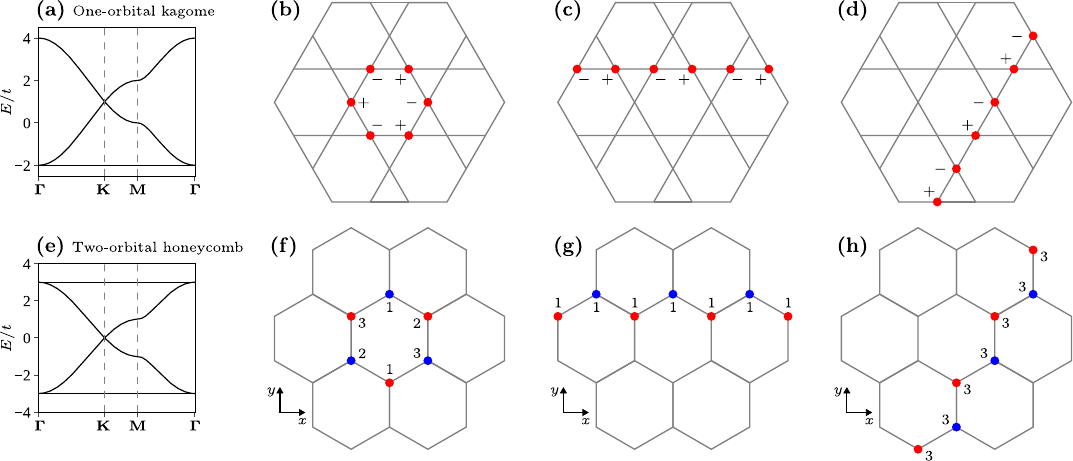}
	\caption{\textbf{(a)} Band structure of the one-orbital kagome lattice Hamiltonian $H_{\text{kag}}$ in \cref{eq:one-orbital-kagome}. \textbf{(b)} Illustration of a prototypical compact localized state for $H_{\text{kag}}$. \textbf{(c)}, \textbf{(d)} Illustrations of two noncontractible loop states for $H_{\text{kag}}$. \textbf{(e)} Band structure of the two-orbital honeycomb lattice Hamiltonian $H_{\text{hc}}$ in \cref{eq:two-orbital-honeycomb}, with $|t_+| = |t_-| = t$. \textbf{(f)} Illustration of a prototypical compact localized state for $H_{\text{hc}}$. \textbf{(g)}, \textbf{(h)} Illustrations of two noncontractible loop states for $H_{\text{hc}}$.}
	\label{appfig:CLS-NLS}
\end{figure}

\subsection{One-orbital kagome model}
We first consider the one-orbital kagome lattice Hamiltonian $H_{\text{kag}}$ in \cref{eq:one-orbital-kagome}. \cref{appfig:CLS-NLS}\textbf{(a)} shows the band structure of $H_{\text{kag}}$, which has one exactly flat band with energy $-2t$. The CLSs and NLSs for this flat band were previously described in Refs. \cite{Bergman2008,Rhim2019}, and we review them now. \cref{appfig:CLS-NLS}\textbf{(b)} illustrates a prototypical CLS, which is supported on a hexagon of sites shown in red. The coefficients are $+1$ and $-1$ in an alternating pattern, with signs shown in the figure. It is straightforward to check that such a state is an eigenstate of $H_{\text{kag}}$ with energy $-2t$. Although there are $N_{\text{tot}}$ such CLSs, the sum of all these states is zero. As a result, there are $N_{\text{tot}} - 1$ linearly independent CLSs.

\cref{appfig:CLS-NLS}\textbf{(c)},\textbf{(d)} illustrate two NLSs, which are supported on the noncontractible loops of sites shown in red. The coefficients are $+1$ and $-1$ in an alternating pattern, with signs shown in the figures. It is straightforward to check that these states are also eigenstates of $H_{\text{kag}}$ with energy $-2t$. The $N_{\text{tot}}-1$ CLSs and two NLSs form a basis for the flat band and its band touching.

\subsection{Two-orbital honeycomb model}
Next, we consider the two-orbital honeycomb lattice Hamiltonian $H_{\text{hc}}$ in \cref{eq:two-orbital-honeycomb}. $H_{\text{hc}}$ has two exactly flat bands when $|t_+| = |t_-|$ \cite{Wu2007,Calugaru2022,Scheer2023}. However, the two cases $t_+ = t_-$ and $t_+ = -t_-$ are related by a sign reversal on the $\ell = -1$ orbitals. As a result, we can choose $t_+ = t_- = t$ without loss of generality. Similarly, changing the axis with respect to which the $\varphi_{j', j}$ angles are defined is equivalent to changing the phases of the orbitals. This implies that we lose no generality by defining $\varphi_{j',j}$ relative to the $y$ axis shown in \cref{appfig:CLS-NLS}\textbf{(f)}-\textbf{(h)}.

We indicate the two sublattices of the honeycomb lattice by blue and red colors in \cref{appfig:CLS-NLS}\textbf{(f)}-\textbf{(h)}. We assign these sublattices the values $\alpha = +$ (blue) and $\alpha = -$ (red) and define $C$ to be a diagonal operator whose entries are these sublattice values. \cref{appfig:CLS-NLS}\textbf{(e)} shows the band structure of $H_{\text{hc}}$, which has two exactly flat bands with energies $-3t$ and $3t$. In fact, the entire band structure is symmetric about energy $0$ due to the anticommutation identity $\{H_{\text{hc}}, C\} = 0$. As a result, it suffices to find the CLSs and NLSs for the flat band with energy $-3t$, since the two flat bands are related by the $C$ operator.

\cref{appfig:CLS-NLS}\textbf{(f)} illustrates a prototypical CLS, which is supported on a hexagon of sites colored blue and red. Each site carries an integer label, and the coefficient for the $\ell$ orbital on a site with label $n$ is $\ell e^{-i\ell\zeta_n}$ where $\zeta_n = \frac{2\pi}{3}(n-1)$. The coefficients do not depend on sublattice. We now check that this CLS is indeed an eigenstate of $H_{\text{hc}}$ with energy $-3t$. To do so, we compute the amplitude on a state $\ket{j', \ell'}$ after $H_{\text{hc}}$ is applied to the CLS. If $j'$ is neither in the CLS nor is a nearest neighbor of a site in the CLS, then the amplitude is zero. Next, suppose that $j'$ is not in the CLS, but is a nearest neighbor of a site in the CLS with label $n$. The amplitude is then
\begin{equation}
t\sum_{\ell=\pm 1} e^{i(\ell - \ell')\zeta_n} \ell e^{-i\ell\zeta_n} = t e^{-i\ell' \zeta_n} \sum_{\ell=\pm 1} \ell = 0.
\end{equation}
Finally, suppose that $j'$ is in the CLS and has label $n'$. The amplitude is then
\begin{equation}
\begin{split}
t \sum_{\ell=\pm 1} \sum_{\substack{1 \leq n \leq 3\\n \neq n'}} e^{i(\ell - \ell')(2\pi - \zeta_n - \zeta_{n'})} \ell e^{-i\ell \zeta_n} &= t \sum_{\substack{1 \leq n \leq 3\\n \neq n'}} \ell' e^{-i\ell' \zeta_n} + e^{-i\ell'(\zeta_n + \zeta_{n'})}(-\ell') e^{i\ell' \zeta_n}\\
&= t \ell' \sum_{\substack{1 \leq n \leq 3\\n \neq n'}} e^{-i\ell' \zeta_n} - e^{-i\ell' \zeta_{n'}}\\
&= -3t \left(\ell' e^{-i\ell' \zeta_{n'}}\right).
\end{split}
\end{equation}
We conclude that the CLS shown in \cref{appfig:CLS-NLS}\textbf{(f)} is indeed an eigenstate of $H_{\text{hc}}$ with energy $-3t$. Although there are $N_{\text{tot}}$ CLSs of the form shown in \cref{appfig:CLS-NLS}\textbf{(f)}, the sum of all these states is zero. As a result, there are $N_{\text{tot}} - 1$ linearly independent CLSs.

\cref{appfig:CLS-NLS}\textbf{(g)},\textbf{(h)} illustrate two NLSs, which are supported on the noncontractible loops of sites colored blue and red. Each site carries an integer label, and in either NLS the coefficient for the $\ell$ orbital on a site of sublattice $\alpha$ and label $n$ is $\alpha \ell e^{-i\ell\zeta_n}$ where $\zeta_n = \frac{2\pi}{3}(n-1)$. We now check that these NLSs are indeed eigenstates of $H_{\text{hc}}$ with energy $-3t$. To do so, we compute the amplitude on a state $\ket{j', \ell'}$ after $H_{\text{hc}}$ is applied to an NLS. If $j'$ is neither in the NLS nor is a nearest neighbor of a site in the NLS, then the amplitude is zero. Next, suppose that $j'$ is not in the NLS, but is a nearest neighbor of a site in the NLS with sublattice $\alpha$ and label $n$. The amplitude is then
\begin{equation}
t\sum_{\ell=\pm 1} e^{i(\ell - \ell')\zeta_n} \alpha\ell e^{-i\ell\zeta_n} = t \alpha e^{-i\ell' \zeta_n} \sum_{\ell=\pm 1} \ell = 0.
\end{equation}
Finally, suppose that $j'$ is in the CLS and has sublattice $\alpha'$ and label $n'$. The amplitude is then
\begin{equation}
\begin{split}
t \sum_{\ell=\pm 1} \sum_{\substack{1 \leq n \leq 3\\n \neq n'}} e^{i(\ell - \ell')\zeta_n} (-\alpha')\ell e^{-i\ell \zeta_{n'}} &= -t\alpha' \sum_{\substack{1 \leq n \leq 3\\n \neq n'}} \ell' e^{-i\ell'\zeta_{n'}} + e^{i\ell'\zeta_n}(-\ell')e^{i\ell'\zeta_{n'}}\\
&= -t\alpha'\ell'\left(2 e^{-i\ell'\zeta_{n'}} - e^{i\ell'\zeta_{n'}}\sum_{\substack{1 \leq n \leq 3\\n \neq n'}} e^{i\ell' \zeta_n} \right)\\
&= -3t(\alpha'\ell' e^{-i\ell'\zeta_{n'}}).
\end{split}
\end{equation}
We conclude that the NLSs shown in \cref{appfig:CLS-NLS}\textbf{(g)},\textbf{(h)} are indeed eigenstates of $H_{\text{hc}}$ with energy $-3t$. The $N_{\text{tot}}-1$ CLSs and two NLSs form a basis for the flat band of energy $-3t$ and its band touching.

\section{Parameter variation}\label{app:parameter-variation}
\begin{figure}
	\centering
	\includegraphics{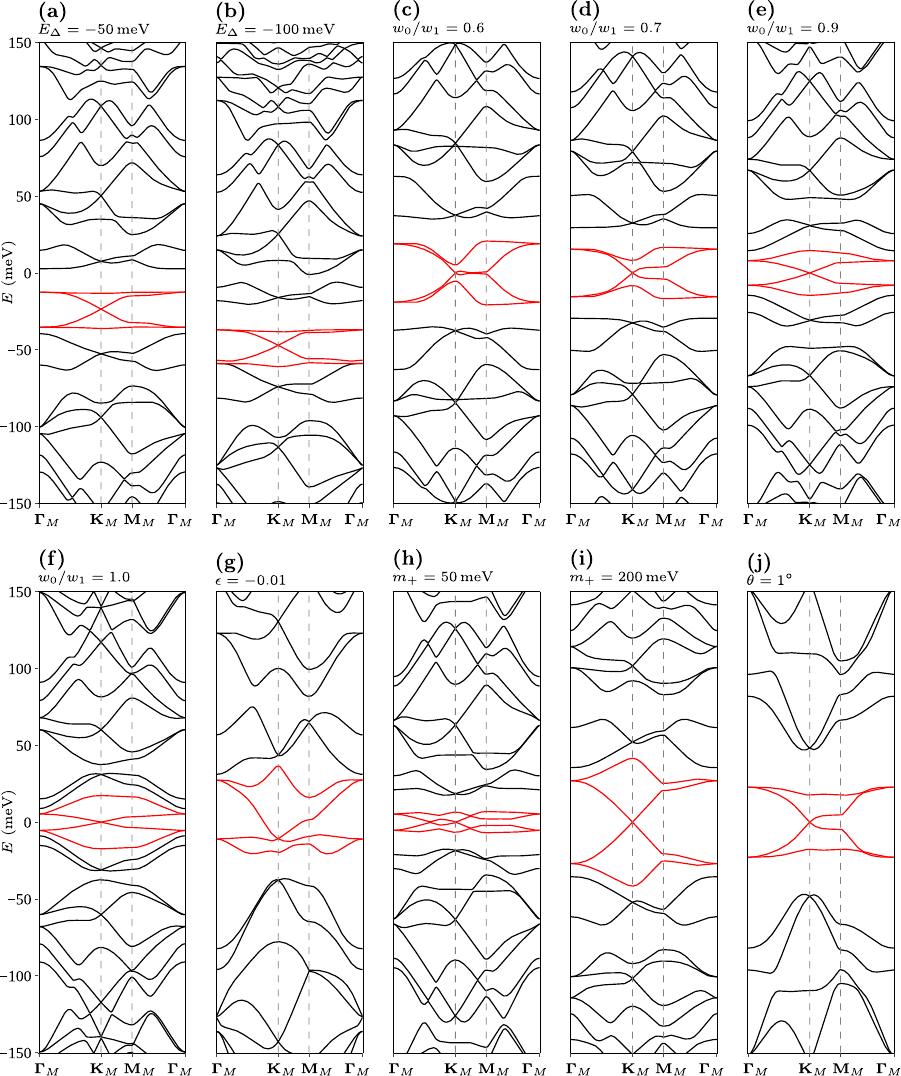}
	\caption{Variations on the TGKG band structure in \cref{fig:bands}\textbf{(a)}. In each panel, all parameters are fixed at those used in \cref{fig:bands}\textbf{(a)} except as indicated. Bands $-2 \leq n \leq 2$ are shown in red. Except in \textbf{(b)}, these bands always form an isolated honeycomb lattice model with EBCR $({^1E}{^2E})_{2b}$ of $P61'$. In \textbf{(b)}, bands $-1 \leq n \leq 2$ and $-4 \leq n \leq -2$ correspond to EBCRs $({^1E_1}{^2E_1})_{1a} \oplus (B)_{1a}$ and $({^1E_2}{^2E_2})_{1a} \oplus (A)_{1a}$ of $P61'$, respectively, and are therefore triangular lattice models.}
	\label{appfig:bands-extra-TGKG}
\end{figure}

\begin{figure}
	\centering
	\includegraphics{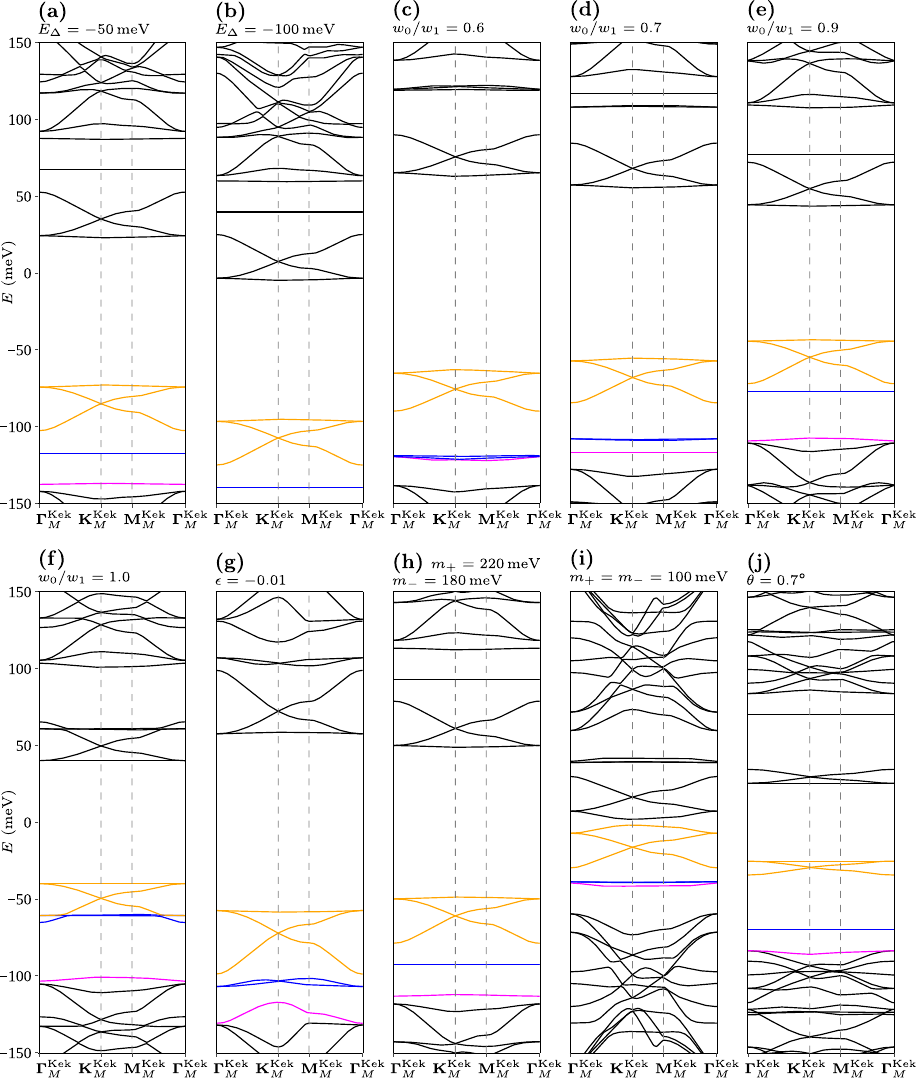}
	\caption{Variations on the TBKG band structure in \cref{fig:bands}\textbf{(b)}. In each panel, all parameters are fixed at those used in \cref{fig:bands}\textbf{(b)} except as indicated. Bands $-3 \leq n \leq -1$ are shown in orange, bands $-5 \leq n \leq -4$ are shown in blue, and band $n = -6$ is shown in magenta. Except in \textbf{(f)}, bands $-3 \leq n \leq -1$ and $1 \leq n \leq 3$ always form isolated kagome lattice flat band models with EBCRs $(B)_{3c}$ and $(A)_{3c}$ of $P61'$, respectively. In \textbf{(f)}, bands $n = \pm 1$ are still extremely flat and bands $-4 \leq n \leq -1$ and $1 \leq n \leq 4$ correspond to EBCRs $(B)_{3c} \oplus ({^1E_2}{^2E_2})_{1a}$ and $(A)_{3c} \oplus ({^1E_1}{^2E_1})_{1a}$ of $P61'$, respectively.}
	\label{appfig:bands-extra-TBKG}
\end{figure}

We now discuss the effects of parameter variation on the band structures of TGKG and TBKG. \cref{appfig:bands-extra-TGKG,appfig:bands-extra-TBKG} show variations on the band structures in \cref{fig:bands}, with one parameter varied in each panel. We see that the structure of the two-orbital honeycomb bands in TGKG is quite sensitive to $w_0/w_1$, $\epsilon$, $m_+$ and $\theta$, but is relatively insensitive to $E_\Delta$. In each case, the two-orbital honeycomb bands remain isolated. On the other hand, the structure of the one-orbital kagome bands in TBKG is quite insensitive to all parameters considered, but the bands are no longer isolated in \cref{appfig:bands-extra-TBKG}\textbf{(f)}.

\begin{figure}
	\centering
	\includegraphics{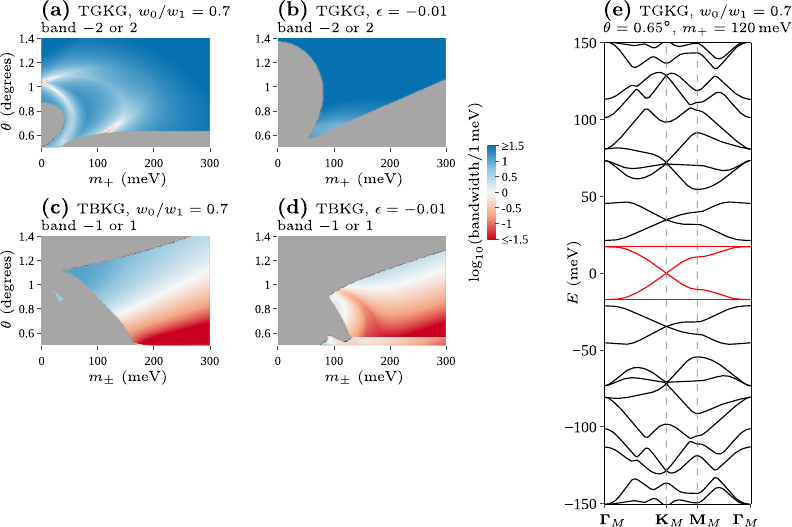}
	\caption{\textbf{(a)}, \textbf{(b)} Variations on the plots in \cref{fig:bandwidths}\textbf{(a)}. \textbf{(c)}, \textbf{(d)} Variations on the plots in \cref{fig:bandwidths}\textbf{(b)}. In each panel, all parameters are fixed at those used in \cref{fig:bandwidths} except for $w_0/w_1$ or $\epsilon$, as indicated. \textbf{(e)} Band structure of TGKG with $w_0/w_1 = 0.7$ and the modified magic parameters $\theta = 0.65^\circ$ and $m_+ = \SI{120}{\milli\electronvolt}$ from \textbf{(a)}.}
	\label{appfig:bandwidths-extra}
\end{figure}

\cref{appfig:bandwidths-extra}\textbf{(a)} shows the bandwidth of the $n = \pm 2$ two-orbital honeycomb lattice flat bands in TGKG as a function of $\theta$ and $m_+$, as in \cref{fig:bandwidths}\textbf{(a)}, except that we now take $w_0/w_1 = 0.7$. We see that upon changing the value of $w_0/w_1$, the magic TGKG regime moves near $\theta = 0.65^\circ$, $m_+ = \SI{120}{\milli\electronvolt}$. \cref{appfig:bandwidths-extra}\textbf{(e)} shows the band structure with these modified magic parameters, which exhibits a two-orbital honeycomb lattice flat band model.

Similarly, \cref{appfig:bandwidths-extra}\textbf{(b)} is identical to \cref{fig:bandwidths}\textbf{(a)}, except that we now take $\epsilon = -0.01$. In this case, we do not observe a magic TGKG regime. \cref{appfig:bandwidths-extra}\textbf{(c)},\textbf{(d)} are identical to \cref{fig:bandwidths}\textbf{(b)} except that $w_0/w_1 = 0.7$ in \cref{appfig:bandwidths-extra}\textbf{(c)} and $\epsilon = -0.01$ in \cref{appfig:bandwidths-extra}\textbf{(d)}. Both of these plots indicate the presence of flat one-orbital kagome bands in a large parameter regime.

It is worth noting that the results in Refs. \cite{Bao2021,Qu2022} which experimentally realize Kekul\'e-O graphene are consistent with $\epsilon = 0$. For a theoretical justification of the small magnitude of $\epsilon$, one can consult DFT calculations of lithium intercalated graphene. For example, the results in Ref. \cite{Wang2014} with density functional vdW-optPBE produce $\epsilon \approx -0.01$.

\end{document}